\documentclass[12pt]{article}
\usepackage{graphicx}

\newcommand\pubnumber{SNSN-323-63}
\newcommand\pubdate{\today}

\def\napoli{Department of Physics, Columbia University, 538 West 120th Street, New York,
NY 10027, USA\\
Department of Physics, University of Houston, Houston, TX 77204, USA
}

\def\Title#1{\begin{center} {\Large #1 } \end{center}}
\def\Author#1{\begin{center}{ \sc #1} \end{center}}
\def\Address#1{\begin{center}{ \it #1} \end{center}}

\newcommand\pubblock{\rightline{\begin{tabular}{l} \pubnumber\\
         \pubdate  \end{tabular}}}
\newenvironment{Abstract}{\begin{quotation}  }{\end{quotation}}
\newenvironment{Presented}{\begin{quotation} \begin{center} 
             PRESENTED AT\end{center}\bigskip 
      \begin{center}\begin{large}}{\end{large}\end{center} \end{quotation}}
\def\Acknowledgements{\bigskip  \bigskip \begin{center} \begin{large}
             \bf ACKNOWLEDGEMENTS \end{large}\end{center}}

\def\beq{\begin{equation}}
\def\eeq#1{\label{#1}\end{equation}}
\def\eeqn{\end{equation}}

\def\beqa{\begin{eqnarray}}
\def\eeqa#1{\label{#1}\end{eqnarray}}
\def\eeqan{\end{eqnarray}}

\let\bar=\overbar

\def\Dslash{\not{\hbox{\kern-4pt $D$}}}
\def\dslash{\not{\hbox{\kern-2pt $\del$}}}

\def\msb{{\bar{\ssstyle M \kern -1pt S}}}

\begin{document}
\begin{titlepage}
\pubblock

\vfill
\Title{Extracting shear viscosity of the Quark Gluon Plasma in the presence of bulk viscosity}
\vfill
\Author{Jacquelyn Noronha-Hostler}
\Address{\napoli}
\vfill
\begin{Abstract}
One of the most remarkable features of the Quark Gluon Plasma is its nearly perfect fluidity behavior indicated by the small shear viscosity to entropy density ratio obtained from fitting relativistic viscous hydrodynamics flow harmonics to experimental data.  In recent years, bulk viscosity has also been considered in the context of event-by-event relativistic hydrodynamics and it has been found to have a non-trivial interplay with shear viscosity. In this paper some of the issues are discussed that require further work when extracting the shear viscosity to entropy density ratio in the presence of a non-zero bulk viscosity. 
\end{Abstract}
\vfill
\begin{Presented}
CIPANP2015\\
Vail, CO USA,  May 19-24, 2015
\end{Presented}
\vfill
\end{titlepage}
\def\thefootnote{\fnsymbol{footnote}}
\setcounter{footnote}{0}

\section{Introduction}

The nearly perfect fluidity of the Quark Gluon Plasma was discovered in ultrarelativistic nucleus-nucleus collisions at RHIC in the early 2000's and it was subsequently confirmed at the LHC at higher collision energies. Initially, the focus was only on the effects of shear viscosity on event-by-event hydrodynamical calculations of flow harmonics \cite{Gale:2012rq}, which are used to study the collective motion of the Quark Gluon Plasma, but in recent years the inclusion of bulk viscosity effects on event-by-event simulations has been shown to affect both the flow harmonics and particle spectra \cite{Noronha-Hostler:2013gga,Noronha-Hostler:2014dqa} while also improving the mapping between the initial conditions and the final flow harmonics \cite{Gardim:2014tya}. More recently, it has been pointed out in \cite{Ryu:2015vwa} that for IP-Glasma initial conditions \cite{Schenke:2012wb} bulk viscosity is needed in order to fit $\langle p_T\rangle$.  

\begin{figure}[htb]
\centering
\begin{tabular}{c c}
\includegraphics[height=2.5in]{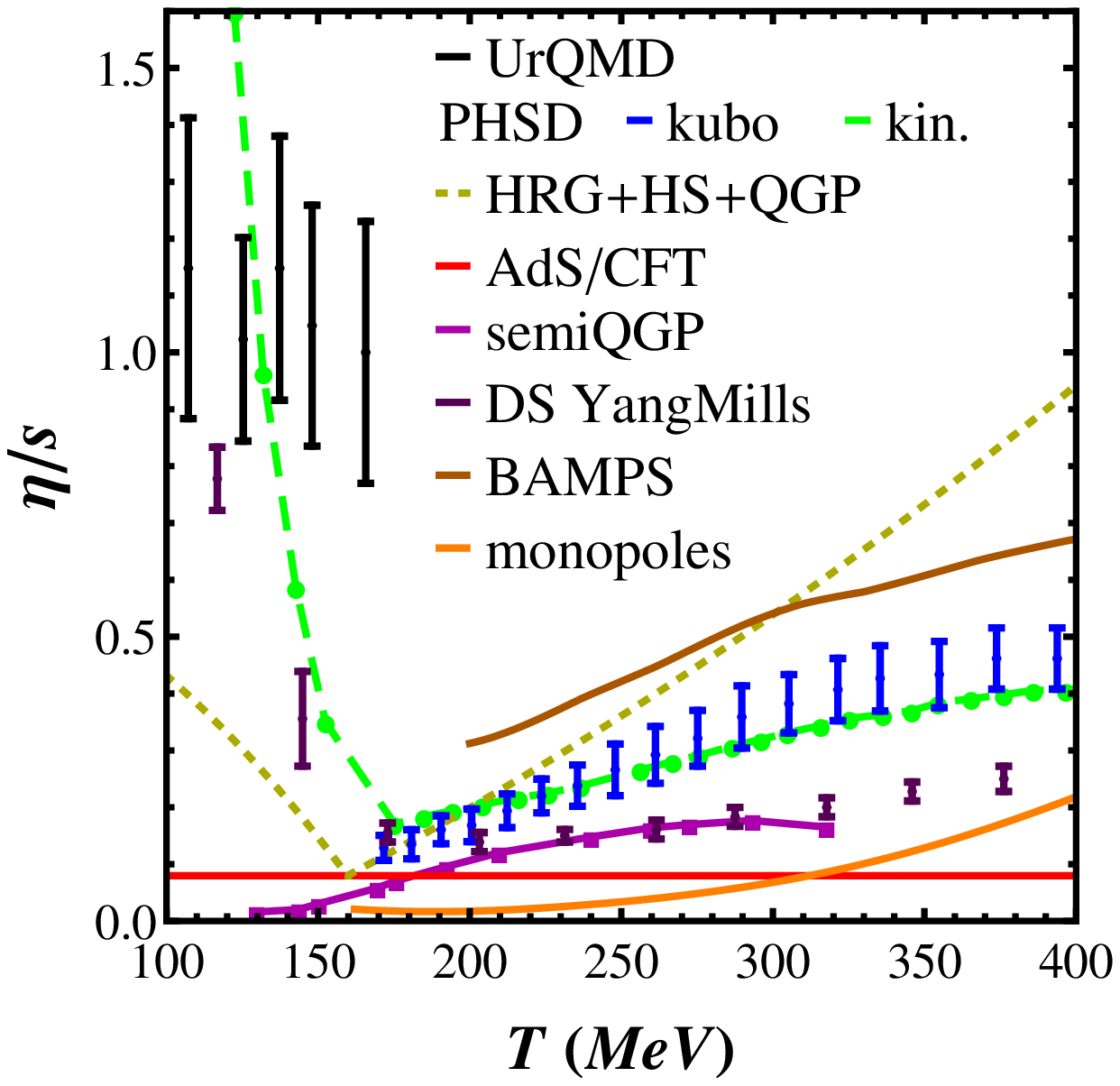} & \includegraphics[height=2.5in]{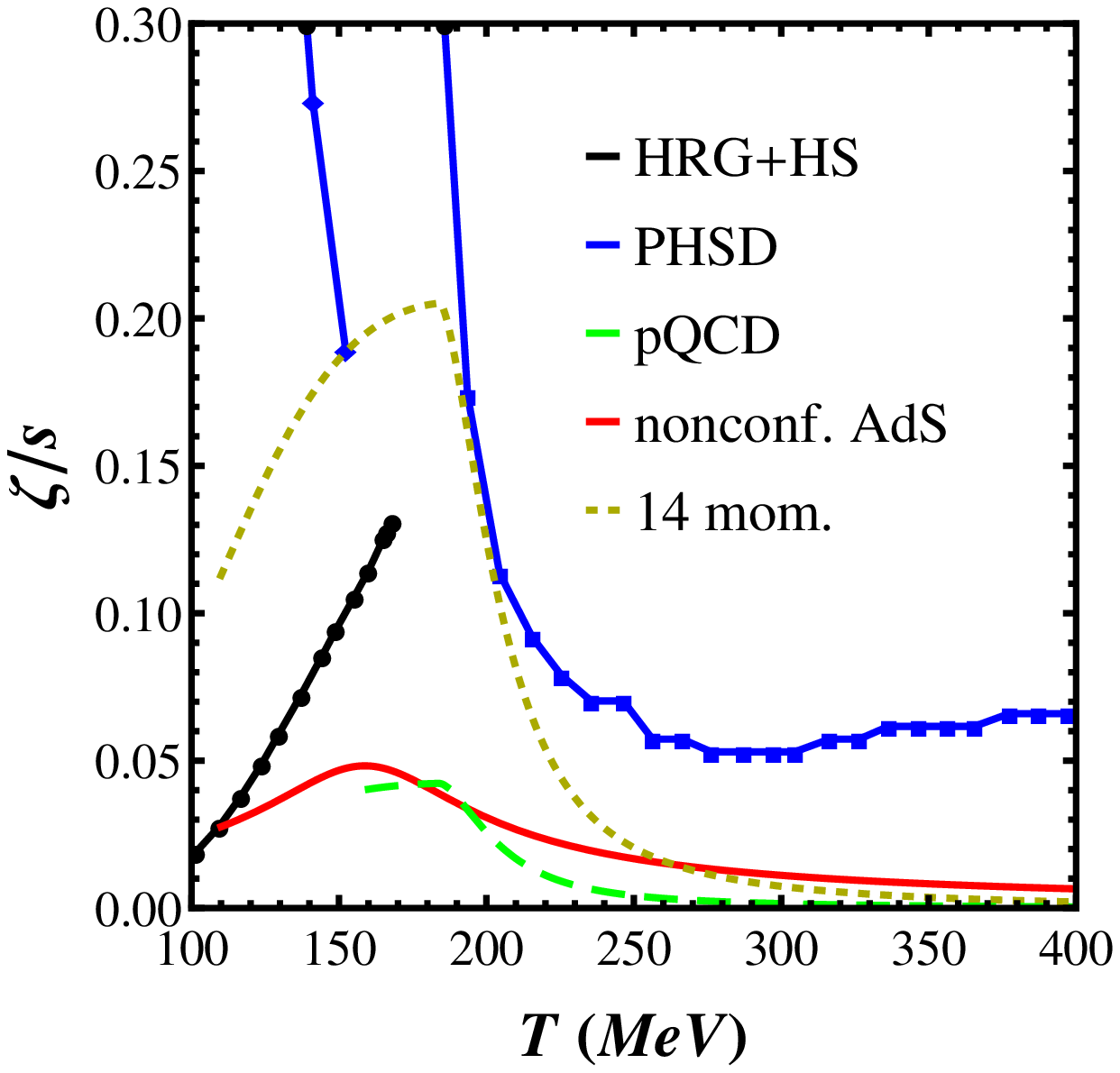}
\end{tabular}
\caption{(Color online) Recent theoretical calculations of $\eta/s(T)$ (left) and $\zeta/s(T)$ (right) are shown for temperatures relevant for heavy ion collisions (the corresponding references for each of the curves can be found in the text).}
\label{fig:Tdep}
\end{figure}

A number of theoretical calculations have been done over the years to uncover the temperature dependence of the shear viscosity to entropy density ratio, $\eta/s(T)$ and the bulk viscosity to entropy density ratio, $\zeta/s(T)$. An overview of those results is shown in Fig.\ \ref{fig:Tdep}. Shear viscosity measures the resistance to gradual deformation by shear stress and the actual effect in hydrodynamics appears from smoothing out the energy density gradients over time.  In Fig.\ \ref{fig:Tdep} (left) the curves corresponds to different calculations of $\eta/s(T)$ coming from a variety of sources such as transport models like URQMD \cite{Demir:2008tr} (in the hadron gas phase), PHSD  \cite{Ozvenchuk:2012kh} (both the hadron gas/QGP phase), and BAMPS \cite{Wesp:2011yy} (QGP phase) or other types of theoretical models such as the gauge/gravity duality \cite{Kovtun:2004de} (for a discussion on temperature dependence in this case see \cite{Cremonini:2012ny}), Hagedorn States \cite{NoronhaHostler:2008ju,NoronhaHostler:2012ug} (see also \cite{Kadam:2014cua}), the semi-QGP model \cite{Hidaka:2009ma}, pure Yang Mills theory \cite{Christiansen:2014ypa}, and color magnetic monopoles \cite{Liao:2008jg,Xu:2014tda}. However, an interesting point to be made is that in order to obtain a minimum in $\eta/s$ around the crossover phase transition region extra degrees of freedom (not present in an ordinary hadron gas) are needed, which can be clearly understood from looking at the entropy density obtained from the Lattice Equation of State \cite{Borsanyi:2013bia} that shows a rapid increase around $T_c \sim 180$ MeV.  These extra degrees of freedom are needed when $T \sim 150-300$ MeV where the system goes from the high temperature end of the hadron gas phase \cite{NoronhaHostler:2008ju,Kadam:2014cua} (modeled using Hagedorn states - massive, short lived resonances that appear only close to the crossover transition) to the strongly interacting Quark Gluon Plasma, which may be modeled by the effects color magnetic monopoles (see \cite{Xu:2014tda}).  

Bulk viscosity acts as a resistance against the volume expansion of a fluid and, as such, it slows down the system's evolution. Bulk viscosity arises in non-conformal systems and it has been suggested \cite{Karsch:2007jc} that $\zeta/s$ may peak near the phase transition induced by the peak in the trace anomaly ($\theta/T^4=(\varepsilon-3p)/T^4$) found on the lattice \cite{Borsanyi:2013bia}. Indeed, models that manage to reproduce the lattice equation of state in the hadron gas phase \cite{NoronhaHostler:2008ju,NoronhaHostler:2012ug} and non-conformal gauge/gravity models \cite{Finazzo:2014cna} are able to produce a peak in the bulk viscosity at the crossover transition.  Other calculations of bulk viscosity include kinetic theory models \cite{Denicol:2014vaa}, weak coupling QCD \cite{Arnold:2006fz}  (see also \cite{Chen:2012jc}), and PHSD \cite{Ozvenchuk:2012kh}. See Fig.\ \ref{fig:Tdep} (right) for a comparison between the models.

Viscous relativistic hydrodynamics can be used as a tool to test the applicability of theoretical calculations of transport coefficients via a comparison to experimental data.  Therefore, the purpose here is to take currently  known $\zeta/s(T)$ theoretical calculations shown in Fig.\ \ref{fig:magnet}  and test if they can fit the flow harmonics but also what size of an effect that they have on the shear viscosity $\eta/s$.  One should note that even parametrizations are based upon theoretical models with important underlying physical assumptions so it is necessary to test these models against experimental data to see the degree of our understanding of transport coefficients.

\section{Setup}

All calculations are done within the 2+1 event-by-event relativistic viscous hydrodynamical code, v-USPhydro, \cite{Noronha-Hostler:2013gga,Noronha-Hostler:2014dqa} that solves the equations of motion of viscous hydrodynamics with bulk and shear viscosity effects using the Lagrangian algorithm Smoothed Particle Hydrodynamics (SPH) which was first adapted for heavy ion collisions in \cite{Aguiar:2000hw}.  For further details on the model, tests, and parameters see \cite{Noronha-Hostler:2013gga,Noronha-Hostler:2014dqa}. NeXus initial conditions are generated using a parton-based Gribov-Regge picture of nucleus-nucleus collisions in which hard partons are treated using perturbative QCD while soft partons are included using the string picture \cite{Drescher:2000ec}, which has been shown to already fit RHIC data well \cite{Gardim:2012yp}. A freeze-out temperature of $T_{FO}=120$ MeV is used and decays are taken into account via an adapted version of the AZHYDRO code \cite{azhydro} that includes hadrons and resonances with masses up to $M=1.7$ GeV.  At freeze-out the distribution function is described as 
\begin{equation}\label{eqn:dis}
f=f_0+\delta f_{shear}+\delta f_{bulk}
\end{equation}
where $f_0$ is the local equilibrium distribution function and $\delta f$ are the out-of-equilibrium contributions associated with shear and bulk viscosity.  The individual components are defined in \cite{Noronha-Hostler:2013gga,Noronha-Hostler:2014dqa} and the $\delta f_{bulk}$ is taken from \cite{Monnai:2010qp}.

\begin{figure}[htb]
\centering
\begin{tabular}{c c}
\includegraphics[height=2in]{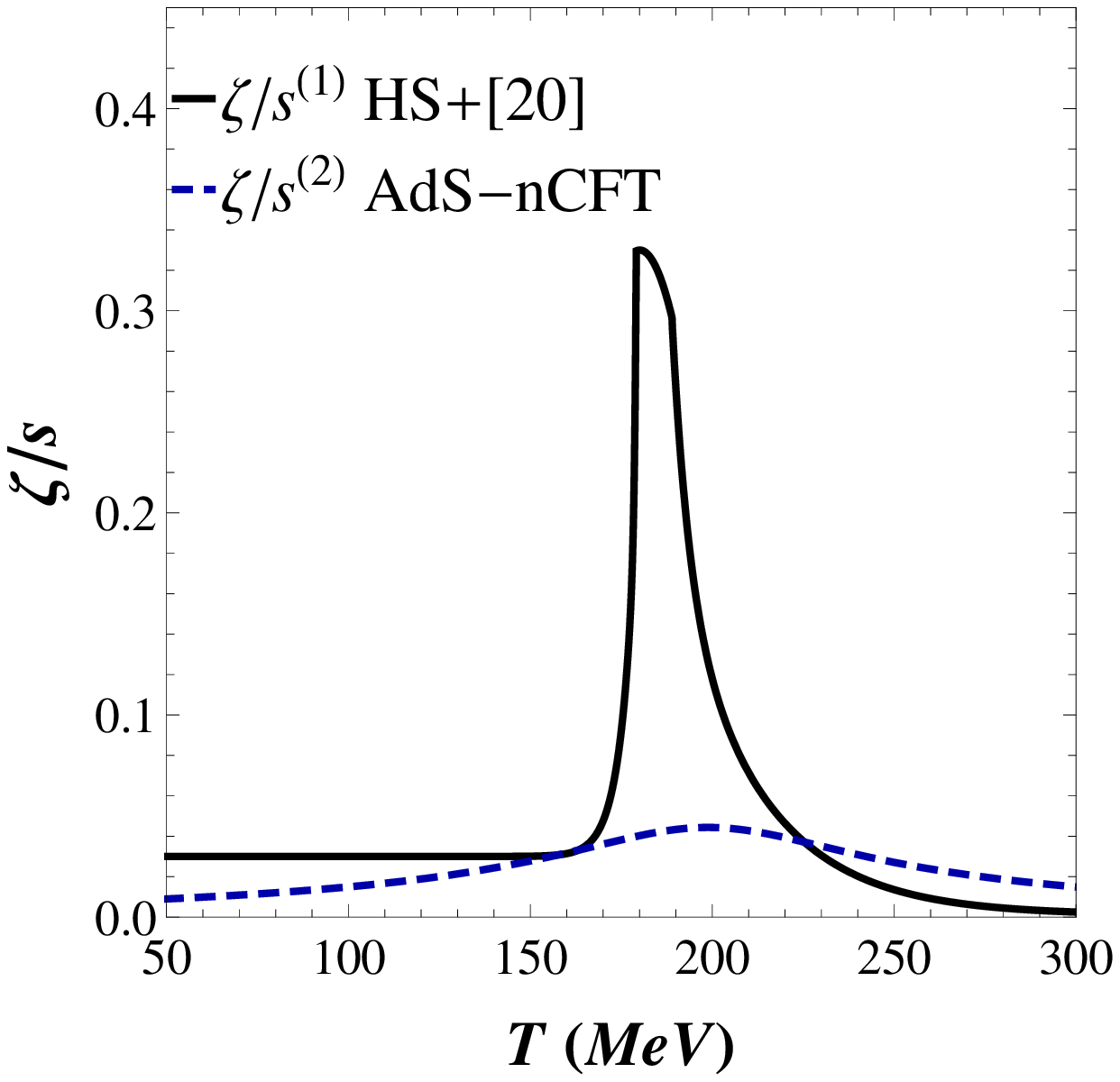} & \includegraphics[height=2in]{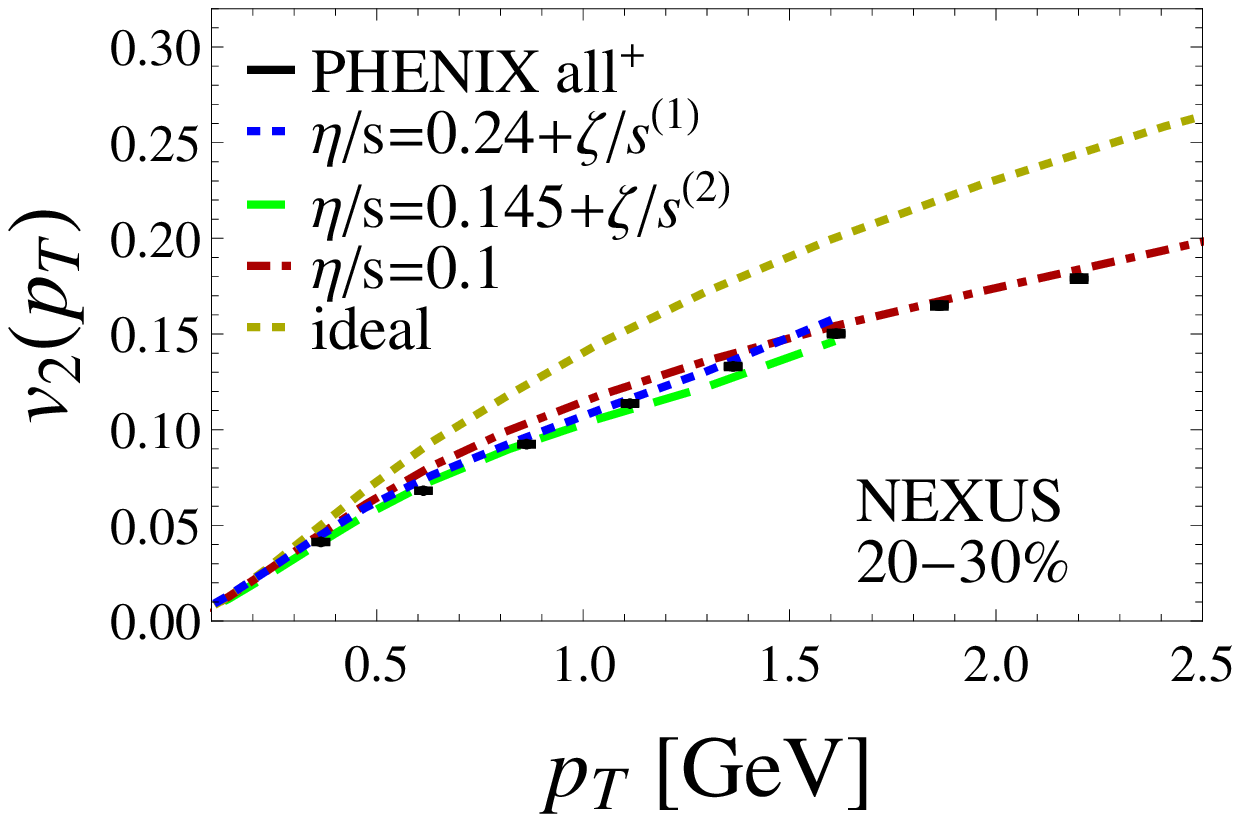}
\end{tabular}
\caption{(Color online) Two theoretical bulk viscosity calculations of $\zeta/s(T)$ (left) and the corresponding hydrodynamical calculation of elliptic flow, $v_2$ at $\sqrt{s}=200$ GeV RHIC collisions for $20-30\%$ centrality (right).}
\label{fig:magnet}
\end{figure}

In Fig.\ \ref{fig:magnet} (left) one can see the two very different choices in $\zeta/s(T)$ taken here.  For $\zeta/s(T)^{(1)}$ \cite{Ryu:2015vwa} the hadron gas phase is taken from a hadron resonance gas model with Hagedorn States that reproduce the lattice Equation of State in \cite{Noronha-Hostler:2013rcw}. Note that in the case of $\zeta/s^{(1)}(T)$, Hagedorn States should be included in the hadronic afterburner for consistency's sake since that generates the large peak in the hadronic sector due to the extra degrees of freedom from these heavy resonances. While this is unlikely to strongly affect the particle ratios \cite{NoronhaHostler:2007jf,NoronhaHostler:2009cf} it will likely affect the differential flow harmonics \cite{Noronha-Hostler:2013rcw}. The curve $\zeta/s(T)^{(2)}$ was generated using the parametrization of the bulk viscosity computed in \cite{Finazzo:2014cna} based on non-conformal holography that also matches the Lattice Equation of State.  As one can see, in both curves $\zeta/s$ has a peak around the phase transition.  However, the peak found in the holography calculation is significantly smaller than the one used in \cite{Ryu:2015vwa}.  

In Fig.\ \ref{fig:magnet} (right) the extraction of $\eta/s=const$ is shown for the two choices in $\zeta/s(T)$.  As a comparison the $v_2$ for shear and ideal hydro calculations are also shown.  Note that in \cite{Gardim:2012yp} higher freeze-out temperatures between $T_{FO}=130-150$ MeV are used. The lines for the inclusion of bulk viscosity are only plotted up to $p_T=1.5$ GeV because beyond that variations in the $\delta f$ contributions are significant \cite{Noronha-Hostler:2013gga,Noronha-Hostler:2014dqa}.  It is clear that when it comes to fitting flow harmonics that the magnitude of shear viscosity is strongly dependent on the choice of bulk viscosity. $\zeta/s(T)^{(1)}$ has a significantly taller peak that is more narrow, which requires a much larger $\eta/s$ to compensate the effect of bulk whereas $\zeta/s(T)^{(2)}$ has a flat, broad peak that only shows a slight increase of $\eta/s$ over the case when only shear viscosity is considered.  However, in both cases the combination of $\zeta/s$ and $\eta/s$ provide a better shape to $v_2$ compared to both ideal and shear.  

One should note that the increase of $\eta/s$ in the presence of bulk is entirely a $\delta f_{bulk}$ effect in Eq.\ (\ref{eqn:dis}).  In \cite{Noronha-Hostler:2013gga} it was shown that as one increases the magnitude of $\zeta/s$ when only the local equilibrium component, $f_0$, is considered then the overall magnitude of integrated $v_n$'s decreases. However, in both \cite{Noronha-Hostler:2013gga,Noronha-Hostler:2014dqa} it was shown that multiple choices of $\delta f_{bulk}$ increase the $v_n$'s, which means that there are two competing effects.  Thus, if one sees a total increase in $v_n$'s in the presence of bulk (or conversely an increase in $\eta/s$ to fit experimental data) this implies that the $\delta f_{bulk}$ ``won" over $f_0$. In \cite{Ryu:2015vwa} the relaxation time approximation was employed to compute $\delta f_{bulk}$, which gives a smaller  contribution so the overall effect was to decrease $v_n$'s as well as decreasing the needed $\eta/s$. This illustrates the necessity for better modeling the non-equilibrium corrections associated with bulk viscosity at freeze-out. 

\begin{table}[t]
\begin{center}
Percentage change of $\langle p_T\rangle$ from $\eta/s$ to $\zeta/s+\eta/s$  \\
\begin{tabular}{l|ccc}  
\hline
Viscosity &  $f_0$ &   $f_0+\delta f_{shear}+\delta f_{bulk}$ \\ \hline
$\zeta/s^{(1)}+\eta/s$  &  $-3.6\%$     &   $-12\%$      \\ 
$\zeta/s^{(2)}+\eta/s$  &  $-2\%$     &    $-1.5\%$      \\ \hline
\end{tabular}
\caption{Percentage change in $\langle p_T\rangle$ of all positively charged particles due to bulk viscosity effects both with and without $\delta f$ corrections compared to shear viscosity only. }
\label{tab:meanpt}
\end{center}
\end{table}

Furthermore, one can also investigate how the magnitude of $\zeta/s$ affects $\langle p_T\rangle$. In Table \ref{tab:meanpt} the percentage change of    $\langle p_T\rangle$ compared to $\eta/s=0.1$ both for only the local equilibrium distribution function $f_0$ and for the inclusion of out-of-equilibrium effects  $f_0+\delta f_{shear}+\delta f_{bulk}$ are shown.  It is clear that bulk viscosity does consistently decrease $\langle p_T\rangle$ even when out-of-equilibrium effects are not considered.  However, in this case the effect is quite small - only $2\%-3\%$ depending on our choice of $\zeta/s$.  In this work the dominant effect behind the decrease in $\langle p_T\rangle$ arises from the out-of-equilibrium corrections to the distribution function.  In the case of $\zeta/s^{(2)}$ because the bulk viscosity is quite small the $\delta f_{shear}$ is the dominant term in the non-equilibrium correction and there is a slight increase of $\langle p_T\rangle$ from $-2\%$ to  $-1.5\%$ with the inclusion of the out-of-equilibrium contributions.  However, for $\zeta/s^{(1)}$ the bulk viscosity is large so $\delta f_{bulk}$ dominates and a large decrease of $\langle p_T\rangle$ from $-3.6\%$  to $-12\%$ is found.  Note that in this paper only NeXus initial conditions are considered whereas the large effect in $\langle p_T\rangle$  in \cite{Ryu:2015vwa} was found using IP-Glasma initial conditions (and a different Ansatz for $\delta f_{bulk}$). One interesting aspect of this is the difference in the overall smoothing scale \cite{Noronha-Hostler:2015coa}, which may be explored in combination with bulk viscosity in the future. 

\section{Conclusions}

In this proceedings the difficulties that arise when extracting the shear viscosity in the presence of bulk viscosity in the context of relativistic hydrodynamic are discussed. Elliptic flow is strongly affected by the choice of the temperature dependence of $\zeta/s$ as well as by the model choice of $\delta f_{bulk}$  corrections to the distribution function at freeze-out.  Assuming the $\delta f_{bulk}$ calculated in \cite{Monnai:2010qp}, the $\eta/s$ can range from 0.1 when $\zeta/s=0$ to $\eta/s=0.24$ in our calculations for $\zeta/s^{(1)}$ that has a peak that reaches $\zeta/s^{(1)}_{max}\approx0.35$.  $\zeta/s^{(2)}$ calculated within the non-conformal holography framework produces a much more modest increase in shear viscosity to only $\eta/s=0.145$.  However, in both cases there is an improvement to the fit to experimental data for NeXus initial conditions.  Additionally, the change in $\langle p_T\rangle$ is also dependent both on the choice of $\zeta/s$ as well as the $\delta f$ correction.  When only the local equilibrium distribution function is considered (i.e., $\delta f=0$), the percentage change of $\langle p_t\rangle$ from the case with only shear viscosity to the one with shear+bulk between the $\zeta/s^{(1)}$ and $\zeta/s^{(2)}$ are quite small, only $-3.6\%$ and $-2\%$, respectively.  However, the largest effect on the $\langle p_T\rangle$ arises from the $\delta f_{bulk}$ correction.  This indicates that if the out-of-equilibrium contribution to the distribution function is large than the $\langle p_T\rangle$ can be a good distinguishing factor in determining $\zeta/s(T)$ that best matches experimental data.  

In conclusion, bulk viscosity has a very complicated interplay with shear viscosity, which indicates that experimentally extracting the exact values of both bulk viscosity and shear viscosity from data will be difficult.  One of the most important issues that remains is determining the correct description of $\delta f_{bulk}$ since it plays a large role in the overall effect of bulk viscosity both for the flow harmonics as well as for $\langle p_T\rangle$, which is not seen when one only includes the equilibrium distribution function. 
Ideally, the field will converge to a comprehensive approach with temperature dependent transport coefficients that are consistent with the lattice equation of state while covering the entire temperature range of a heavy-ion collisions, which would include all needed hadronic resonances to reproduce these transport coefficients in the hadron resonance gas phase as well.  As more energies are explored at RHIC and the LHC  the hope is that more constraints may be placed on the transport coefficients.

\Acknowledgements
I would like to thank F.~Gardim for providing NeXus initial conditions. This work was supported in part by the US-DOE Nuclear Science Grant No. DE-FG02-93ER40764.


\begin{thebibliography}{99}

%
%
  
\bibitem{Gale:2012rq} 
  C.~Gale, S.~Jeon, B.~Schenke, P.~Tribedy and R.~Venugopalan,
  Phys.\ Rev.\ Lett.\  {\bf 110}, no. 1, 012302 (2013)
  doi:10.1103/PhysRevLett.110.012302
  [arXiv:1209.6330 [nucl-th]].
  
\bibitem{Noronha-Hostler:2013gga} 
  J.~Noronha-Hostler, G.~S.~Denicol, J.~Noronha, R.~P.~G.~Andrade and F.~Grassi,
  Phys.\ Rev.\ C {\bf 88}, 044916 (2013)
  [arXiv:1305.1981 [nucl-th]].
  
\bibitem{Noronha-Hostler:2014dqa} 
  J.~Noronha-Hostler, J.~Noronha and F.~Grassi,
  Phys.\ Rev.\ C {\bf 90}, no. 3, 034907 (2014)
  [arXiv:1406.3333 [nucl-th]].
  
  
\bibitem{Gardim:2014tya} 
  F.~G.~Gardim, J.~Noronha-Hostler, M.~Luzum and F.~Grassi,
  Phys.\ Rev.\ C {\bf 91}, no. 3, 034902 (2015)
  doi:10.1103/PhysRevC.91.034902
  [arXiv:1411.2574 [nucl-th]].
  
\bibitem{Ryu:2015vwa} 
  S.~Ryu, J.-F.~Paquet, C.~Shen, G.~S.~Denicol, B.~Schenke, S.~Jeon and C.~Gale,
  arXiv:1502.01675 [nucl-th].
  
\bibitem{Schenke:2012wb} 
  B.~Schenke, P.~Tribedy and R.~Venugopalan,
  Phys.\ Rev.\ Lett.\  {\bf 108}, 252301 (2012)
  doi:10.1103/PhysRevLett.108.252301
  [arXiv:1202.6646 [nucl-th]].
  
  
\bibitem{Demir:2008tr} 
  N.~Demir and S.~A.~Bass,
  Phys.\ Rev.\ Lett.\  {\bf 102}, 172302 (2009)
  doi:10.1103/PhysRevLett.102.172302
  [arXiv:0812.2422 [nucl-th]].
  
\bibitem{Ozvenchuk:2012kh} 
  V.~Ozvenchuk, O.~Linnyk, M.~I.~Gorenstein, E.~L.~Bratkovskaya and W.~Cassing,
  Phys.\ Rev.\ C {\bf 87}, no. 6, 064903 (2013)
  doi:10.1103/PhysRevC.87.064903
  [arXiv:1212.5393 [hep-ph]].
  
\bibitem{Wesp:2011yy} 
  C.~Wesp, A.~El, F.~Reining, Z.~Xu, I.~Bouras and C.~Greiner,
  Phys.\ Rev.\ C {\bf 84}, 054911 (2011)
  doi:10.1103/PhysRevC.84.054911
  [arXiv:1106.4306 [hep-ph]].
  
    
\bibitem{Kovtun:2004de} 
  P.~Kovtun, D.~T.~Son and A.~O.~Starinets,
  Phys.\ Rev.\ Lett.\  {\bf 94}, 111601 (2005)
  doi:10.1103/PhysRevLett.94.111601
  [hep-th/0405231].


\bibitem{Cremonini:2012ny} 
  S.~Cremonini, U.~Gursoy and P.~Szepietowski,
  JHEP {\bf 1208}, 167 (2012)
  doi:10.1007/JHEP08(2012)167
  [arXiv:1206.3581 [hep-th]].
  
\bibitem{NoronhaHostler:2008ju} 
  J.~Noronha-Hostler, J.~Noronha and C.~Greiner,
  Phys.\ Rev.\ Lett.\  {\bf 103}, 172302 (2009)
  [arXiv:0811.1571 [nucl-th]].
  
\bibitem{NoronhaHostler:2012ug} 
  J.~Noronha-Hostler, J.~Noronha and C.~Greiner,
  Phys.\ Rev.\ C {\bf 86}, 024913 (2012)
  doi:10.1103/PhysRevC.86.024913
  [arXiv:1206.5138 [nucl-th]].
  


\bibitem{Kadam:2014cua} 
  G.~P.~Kadam and H.~Mishra,
  Nucl.\ Phys.\ A {\bf 934}, 133 (2014)
  doi:10.1016/j.nuclphysa.2014.12.004
  [arXiv:1408.6329 [hep-ph]].



\bibitem{Hidaka:2009ma} 
  Y.~Hidaka and R.~D.~Pisarski,
  Phys.\ Rev.\ D {\bf 81}, 076002 (2010)
  doi:10.1103/PhysRevD.81.076002
  [arXiv:0912.0940 [hep-ph]].
  
  
\bibitem{Christiansen:2014ypa} 
  N.~Christiansen, M.~Haas, J.~M.~Pawlowski and N.~Strodthoff,
  Phys.\ Rev.\ Lett.\  {\bf 115}, no. 11, 112002 (2015)
  doi:10.1103/PhysRevLett.115.112002
  [arXiv:1411.7986 [hep-ph]].
  
\bibitem{Liao:2008jg} 
  J.~Liao and E.~Shuryak,
  Phys.\ Rev.\ Lett.\  {\bf 101}, 162302 (2008)
  doi:10.1103/PhysRevLett.101.162302
  [arXiv:0804.0255 [hep-ph]].
  
\bibitem{Xu:2014tda} 
  J.~Xu, J.~Liao and M.~Gyulassy,
  Chin.\ Phys.\ Lett.\  {\bf 32}, no. 9, 092501 (2015)
  doi:10.1088/0256-307X/32/9/092501
  [arXiv:1411.3673 [hep-ph]].
  
\bibitem{Borsanyi:2013bia} 
  S.~Borsanyi, Z.~Fodor, C.~Hoelbling, S.~D.~Katz, S.~Krieg and K.~K.~Szabo,
  Phys.\ Lett.\ B {\bf 730}, 99 (2014)
  doi:10.1016/j.physletb.2014.01.007
  [arXiv:1309.5258 [hep-lat]].

\bibitem{Karsch:2007jc} 
  F.~Karsch, D.~Kharzeev and K.~Tuchin,
  Phys.\ Lett.\ B {\bf 663}, 217 (2008)
  doi:10.1016/j.physletb.2008.01.080
  [arXiv:0711.0914 [hep-ph]].


\bibitem{Finazzo:2014cna} 
  S.~I.~Finazzo, R.~Rougemont, H.~Marrochio and J.~Noronha,
  JHEP {\bf 1502}, 051 (2015)
  [arXiv:1412.2968 [hep-ph]].



\bibitem{Denicol:2014vaa} 
  G.~S.~Denicol, S.~Jeon and C.~Gale,
  Phys.\ Rev.\ C {\bf 90}, no. 2, 024912 (2014)
  [arXiv:1403.0962 [nucl-th]].
  
\bibitem{Arnold:2006fz} 
  P.~B.~Arnold, C.~Dogan and G.~D.~Moore,
  Phys.\ Rev.\ D {\bf 74}, 085021 (2006)
  doi:10.1103/PhysRevD.74.085021
  [hep-ph/0608012].
  
  
  
\bibitem{Chen:2012jc} 
  J.~W.~Chen, Y.~F.~Liu, Y.~K.~Song and Q.~Wang,
  Phys.\ Rev.\ D {\bf 87}, no. 3, 036002 (2013)
  doi:10.1103/PhysRevD.87.036002
  [arXiv:1212.5308 [hep-ph]].
  
%


\bibitem{Aguiar:2000hw} 
  C.~E.~Aguiar, T.~Kodama, T.~Osada and Y.~Hama,
  J.\ Phys.\ G {\bf 27}, 75 (2001)
  doi:10.1088/0954-3899/27/1/306
  [hep-ph/0006239].



\bibitem{Drescher:2000ec} 
  H.~J.~Drescher, S.~Ostapchenko, T.~Pierog and K.~Werner,
  Phys.\ Rev.\ C {\bf 65}, 054902 (2002)
  [hep-ph/0011219].
  
\bibitem{Gardim:2012yp}
  F.~G.~Gardim, F.~Grassi, M.~Luzum and J.~Y.~Ollitrault,
  Phys.\ Rev.\ Lett.\  {\bf 109} (2012) 202302
  doi:10.1103/PhysRevLett.109.202302
  [arXiv:1203.2882 [nucl-th]].
  
\bibitem{azhydro} P.~F.~Kolb, J.~Sollfrank, and U.~Heinz, Phys.\ Rev.\ C {\bf 62} (2000) 054909; P.~F.~Kolb and R.~Rapp, Phys.\ Rev.\ C {\bf 67} (2003) 044903; P.~F.~Kolb and U.~Heinz, nucl-th/0305084. 

\bibitem{Monnai:2010qp} 
  A.~Monnai and T.~Hirano,
  Nucl.\ Phys.\ A {\bf 847}, 283 (2010)
  [arXiv:1003.3087 [nucl-th]].
  
\bibitem{Noronha-Hostler:2013rcw} 
  J.~Noronha-Hostler, J.~Noronha, G.~S.~Denicol, R.~P.~G.~Andrade, F.~Grassi and C.~Greiner,
  Phys.\ Rev.\ C {\bf 89}, no. 5, 054904 (2014)
  doi:10.1103/PhysRevC.89.054904
  [arXiv:1302.7038 [nucl-th]].






\bibitem{NoronhaHostler:2007jf} 
  J.~Noronha-Hostler, C.~Greiner and I.~A.~Shovkovy,
  Phys.\ Rev.\ Lett.\  {\bf 100}, 252301 (2008)
  doi:10.1103/PhysRevLett.100.252301
  [arXiv:0711.0930 [nucl-th]].


\bibitem{NoronhaHostler:2009cf} 
  J.~Noronha-Hostler, M.~Beitel, C.~Greiner and I.~Shovkovy,
  Phys.\ Rev.\ C {\bf 81}, 054909 (2010)
  doi:10.1103/PhysRevC.81.054909
  [arXiv:0909.2908 [nucl-th]].



  


\bibitem{Noronha-Hostler:2015coa} 
  J.~Noronha-Hostler, J.~Noronha and M.~Gyulassy,
  arXiv:1508.02455 [nucl-th].
  
  
  
\bibitem{Noronha-Hostler:2015uye} 
  J.~Noronha-Hostler, M.~Luzum and J.~Y.~Ollitrault,
  arXiv:1511.06289 [nucl-th].
  
%
%


\end{thebibliography}
\end{document}